\documentclass[twocolumn]{autart}

\usepackage[utf8]{inputenc}
\usepackage[T1]{fontenc}
\usepackage{amsmath}
\usepackage{amssymb}
\usepackage{enumerate}
\usepackage{graphicx}
\usepackage{dsfont}
\usepackage{cite}
\usepackage{xcolor}

\renewcommand{\qed}{\relax\ifmmode\hfill\square\else\hfill$\square$\fi}

\makeatletter
\def\ps@copyright{\let\@mkboth\@gobbletwo
  \def\@oddhead{}%
  \let\@evenhead\@oddhead
  \def\@oddfoot{\small\slshape
    \def\@tempa{0}
      Preprint available on arXiv\hfil\@date\/}	
  \let\@evenfoot\@oddfoot
}
\makeatother

\begin{document}

\begin{frontmatter}
    \title{Stabilizing Tube-Based Model Predictive Control:\\ Terminal Set and Cost Construction for LPV Systems\\(extended version)\thanksref{footnoteinfo}}

    \thanks[footnoteinfo]{This work was supported by the Impulse 1 program of the Eindhoven University of Technology and ASML. The material in this paper was not presented at any conference.\\
        * Corresponding author J. Hanema.\\
        \copyright\ 2017. This manuscript version is made available under the CC-BY-NC-ND 4.0 license \texttt{http://creativecommons.org/licenses/by-nc-nd/4.0/}}
    
    \author[csgroup]{J. Hanema}\ead{j.hanema@tue.nl}$^{,\ast}$,
    \author[csgroup]{M. Lazar}\ead{m.lazar@tue.nl},
    \author[csgroup]{R. Tóth}\ead{r.toth@tue.nl}
    
    \address[csgroup]{Department of Electrical Engineering, Eindhoven University of Technology, Eindhoven, The Netherlands.}                                
    
    
    \begin{abstract}
        This paper presents a stabilizing tube-based MPC synthesis for LPV systems.
        We employ terminal constraint sets which are required to be controlled periodically contractive.
        Periodically (or finite-step) contractive sets are easier to compute and can be of lower complexity than ``true'' contractive ones,
        lowering the required computational effort both off-line and on-line.
        Under certain assumptions on the tube parameterization, recursive feasibility of the scheme is proven.
        Subsequently, asymptotic stability of the origin is guaranteed through the construction of a suitable terminal cost based on a novel Lyapunov-like metric for compact convex sets containing the origin.
        A periodic variant on the well-known homothetic tube parameterization that satisfies the necessary assumptions and yields a tractable LPV MPC algorithm is derived.
        The resulting MPC algorithm requires the on-line solution of a single linear program with linear complexity in the prediction horizon.
        The properties of the approach are demonstrated by a numerical example.
    \end{abstract}
\end{frontmatter}

% ==============================================================================

\section{Introduction}

In a \emph{linear parameter-varying} (LPV) system, the state transition map is linear, but this linear map depends on an external \emph{scheduling variable} denoted by $\theta$.
The present work considers systems represented in the state-space form $x(k+1)=A(\theta(k))x(k)+Bu(k)$.
In this setting, the current value $\theta(k)$ can be measured for all time $k$, but the future behavior of $\theta$ is generally not known exactly.
This uncertainty complicates the application of \emph{model predictive control} (MPC),
as guaranteeing recursive feasibility and closed-loop stability necessitates the use of an MPC which is ``robust'' against all possible future scheduling variations.
Predictive control under uncertainty gives rise to a so-called feedback min-max optimization problem \cite{Lee1997,Bemporad2003}, which can be solved, e.g., by dynamic programming.
Because of its inherent complexity, it is useful to search for more conservative, but efficient, approximations of this difficult problem which are computationally tractable in practice \cite{Rakovic2015a}.

\emph{Tube MPC} (TMPC) is a paradigm devised to reduce complexity with respect to the min-max solution.
A principal advantage is that its computational complexity scales well (often, linearly) in the length of prediction horizon.
Tube-based approaches were originally proposed to control constrained linear systems subject to \emph{additive} disturbances \cite{Langson2004,Mayne2005,Rakovic2012a,Rakovic2012,Rakovic2016a}.
In this type of TMPC, the purpose of constructing a tube is to keep the perturbed system trajectories close to a nominal trajectory.
The work \cite{Mayne2005} uses \emph{rigid} tubes, consisting of a sequence of translated copies of a pre-designed robustly positively invariant basic shape set, and ensures recursive feasibility by tightening constraints.
In \cite{Langson2004,Rakovic2012a}, so-called \emph{homothetic} tubes are introduced:
these tubes consist of a sequence of translated and, additionally, scaled copies of the basic shape set.
The scalings are optimized on-line, so no constraint tightening is necessary.
A much more flexible, but more expensive, parameterization is given in \cite{Rakovic2012}, which drops the restriction that the basic shape sets are designed off-line.
The \emph{elastic} tubes of \cite{Rakovic2016a} also improve upon the flexibility of homothetic tubes, by scaling each hyperplane of the basic shape set individually.
In all the TMPC approaches mentioned so far, persistent additive disturbances are assumed to be present, and hence asymptotic stability of the origin can not be established.
Instead, convergence to some limit set centered around the origin is attained.
In contrast, for an LPV system, it should be possible to asymptotically stabilize the origin because the uncertainty enters \emph{multiplicatively} in the state transition matrix.
Thus, the notion of robust stability required for LPV systems is different from the one employed in the aforementioned TMPC approaches for additively disturbed systems.

Besides robust control, tubes can also be applied for the purpose of stabilization.
A framework for the construction of ``stabilizing'' tubes, with application to the predictive control of linear systems on assigned initial condition sets, was presented in \cite{Brunner2013}.
All system trajectories emanating from the initial condition set are restricted to be inside a tube, which terminates in a controlled $\lambda$-contractive terminal set.
The specific form of tube used in \cite{Brunner2013} is homothetic to the terminal set, and in that sense similar to the tubes from \cite{Langson2004,Rakovic2012a}.
However, the purpose of the tube (stabilizing the origin, versus a robustly invariant set \emph{around} the origin) is markedly different.
The theoretical conditions in \cite{Brunner2013} are in principle not limited to the homothetic case, but would then require a new terminal cost design.

The authors of \cite{Langson2004} discuss a possible adaptation of ``additive'' TMPC to parametrically uncertain systems, however without investigating closed-loop stability.
Specialized TMPC approaches for multiplicatively uncertain systems are \cite{Munoz-Carpintero2013a,Fleming2015}:
in these works, there is no scheduling variable that can be measured on-line, classifying them as ``robust'' rather than ``LPV'' approaches.
An LPV tube MPC based on the ``stabilizing tube'' setting of \cite{Brunner2013} was presented in \cite{Hanema2016:cdc:final}:
therein, the constructed tubes are homothetic to a $\lambda$-contractive terminal set, and the on-line optimization of predicted feedback policies is carried out over vertex controls.
The cross sections in \cite{Munoz-Carpintero2013a,Fleming2015} are more flexible than in \cite{Hanema2016:cdc:final}, as each hyperplane of the basic shape set is scaled individually.
From a different perspective, the tubes of \cite{Munoz-Carpintero2013a,Fleming2015} are more restricted because the considered feedback policy consists of control actions superimposed upon a pre-determined linear state feedback, and the tube center has to satisfy a ``nominal'' trajectory.

In this paper, we adopt the basic setting of \cite{Hanema2016:cdc:final}, which evolved from \cite{Brunner2013}.
An important issue in these works is that particularly in the LPV case, the $\lambda$-contractive sets required for stability can be hard to compute and they can be of very high complexity as the state dimension increases.
The main contribution of the current paper is a terminal cost computation based on periodic LPV set dynamics, which allows the construction of stabilizing tube-based predictive controllers for LPV systems using finite-step (or ``periodic'') contractive terminal sets.
Such sets can be viewed as a relaxation of $\lambda$-contractive sets, and are often easier to compute.
The resulting MPC algorithm requires the on-line solution of a single linear program (LP) with a number of variables and constraints which grows linearly in the prediction horizon.

The paper is organized as follows.
In Section 2, we discuss notation, the problem setup, and present the main concepts of finite-step contractive sets.
The general formulation of TMPC with finite-step terminal conditions is given in Section 3.
Suitable parameterizations to enable efficient implementation are presented in Section 4.
Finally, in Section 5, the method is demonstrated on a numerical example.

In this extended paper, the proofs of all lemmas, propositions and theorems can be found in the Appendix.

% ==============================================================================

\section{Preliminaries}

% ------------------------------------------------------------------------------

\subsection{Notation and basic definitions}

The set of nonnegative real numbers is denoted by $\mathbb{R}_+$ and $\mathbb{N}$ denotes the set of nonnegative integers including zero. Define the index set $\mathbb{N}_{[a,b]}$ with $0\leq a\leq b$ as $\mathbb{N}_{[a,b]}:=\{i\in\mathbb{N}\mid a\leq i\leq b\}$.
The predicted value of a variable $z$ at time instant $k+i$ given the information available at time $k$ is denoted by $z_{i|k}$.
In this paper, the notation $\|x\|$ always refers to the $\infty$-norm of a vector $x\in\mathbb{R}^n$, i.e., $\|x\|=\|x\|_\infty=\max_{i\in\mathbb{N}_{[1,n]}} |x_i|$.
Let $\mathcal{C}^n$ denote the set of all compact convex subsets of $\mathbb{R}^n$.
A set $X\in\mathcal{C}^{n}$ which contains the origin in its non-empty interior is called a proper C-set, or PC-set.
The convex hull of a set $X\subset\mathbb{R}^n$ is denoted by $\mathrm{convh}\{X\}$.
A subset of $\mathbb{R}^{n}$ is a polyhedron if it is an intersection of finitely many half-spaces.
A polytope is a compact polyhedron and can equivalently be represented as the convex hull of finitely many points in $\mathbb{R}^n$.
For sets $Y,Z\subset\mathbb{R}^{n}$ and a scalar $\alpha\in\mathbb{R}$ let $\alpha Y=\{\alpha y\mid y\in Y\}$.
Minkowski set addition is defined as $Y\oplus Z =\{y+z\mid y\in Y, z\in Z\}$ and for a vector $v\in\mathbb{R}^{n}$ let $v\oplus Y:=\{v\}\oplus Y$.
The Hausdorff distance between a nonempty set $X\subset\mathbb{R}^{n}$ and the origin is $d_H^0(X)=d_H\left(X,\{0\}\right)=\sup_{x\in X} \|x\|$.
For a vector $x\in\mathbb{R}^{n}$, observe that $d_H^0\left(\{x\}\right) = \|x\|$.
A function $f:\mathbb{R}_+\rightarrow\mathbb{R}_+$ is of class $\mathcal{K}_\infty$ when it is continuous, strictly increasing, $f(0)=0$, and $\lim_{\xi\rightarrow\infty}f(\xi)=\infty$.

The gauge function $\psi_S:\mathbb{R}^n\rightarrow\mathbb{R}_+$ of a given PC-set $S\subset\mathbb{R}^n$ is $\psi_S(x) = \inf\left\{\gamma\mid x\in\gamma S\right\}$ \cite{Schneider2013c1}.
We introduce a generalized ``set''-gauge function.

\begin{defn}\label{def:set_gauge}
    The set-gauge function
    $\Psi_S:\mathcal{C}^n\rightarrow \mathbb{R}_+$ corresponding to a PC-set $S\subset \mathbb{R}^n$ is
    \[\Psi_S(X) := \sup_{x\in X} \psi_S(x) = \inf\left\{\gamma\mid X\subseteq \gamma S\right\}.\]
\end{defn}

The functions $\psi_S(\cdot)$ and $\Psi_S(\cdot)$ are $\mathcal{K}_\infty$-bounded.

\begin{lem}\label{lem:gauge_kinf}
    Let $S\subset \mathbb{R}^n$ be a PC-set. Then, the following properties hold:
    \begin{enumerate}[(i)]
        \item $\exists s_1,s_2\in\mathcal{K}_\infty$ such that $\forall x\in\mathbb{R}^n:\ s_1\left(\|x\|\right) \leq \psi_S(x) \leq s_2\left(\|x\|\right)$,
        \item $\exists s_3,s_4\in\mathcal{K}_\infty$ such that $\forall X\in\mathcal{C}^n:\ s_3\left(d_H^0(X)\right) \leq \Psi_S(X) \leq s_4\left(d_H^0(X)\right)$.
    \end{enumerate}
\end{lem}

% ------------------------------------------------------------------------------

\subsection{Problem Setup}

We consider a constrained LPV system, represented by the following state-space equation
\begin{equation}\label{eq:sys}
    x(k+1) = A(\theta(k))x(k) + Bu(k),\ k\in\mathbb{N},
\end{equation}
with $x(0)=x_0$, and where $u:\mathbb{N}\rightarrow\mathbb{U}\subseteq\mathbb{R}^{n_\mathrm{u}}$ is the input, $x:\mathbb{N}\rightarrow\mathbb{X}\subseteq\mathbb{R}^{n_{\mathrm{x}}}$ is the state variable, and $\theta:\mathbb{N}\rightarrow\Theta\subseteq\mathbb{R}^{n_{\mathrm{\theta}}}$ is the scheduling signal.
The sets $\mathbb{U}$ and $\mathbb{X}$ are the input- and state constraint sets, respectively, while $\Theta$ is called the scheduling set.
The matrix $A(\theta)$ in (\ref{eq:sys}) is assumed to be a real affine function of $\theta$, i.e.,
\begin{equation}\label{eq:def_ab}
    A(\theta) = A_0 + \sum_{i=1}^{n_\theta} \theta_i A_i.
\end{equation}
We consider systems with a constant $B$-matrix, because then all resulting optimization problems will turn out to be convex.
It is, however, possible to transform any system with a parameter-varying $B$ into the form \eqref{eq:sys} by including a stable input filter or a pre-integrator \cite{Blanchini2007}.
The system \eqref{eq:sys} satisfies the following assumptions.

\begin{assum}\label{ass:system_assumptions}
    (i) The values $x(k)$ and $\theta(k)$ can be measured at every time $k\in\mathbb{N}$.
    (ii) The system represented by (\ref{eq:sys}) is stabilizable under the constraints $(\mathbb{X},\mathbb{U})$.
    (iii) The sets $\mathbb{X}$ and $\mathbb{U}$ are polytopic PC-sets.
    (iv) The set $\Theta$ is a polytope with $q$ vertices, i.e., $\Theta=\mathrm{convh}\{\bar{\theta}^j \mid j \in\mathbb{N}_{[1,q]}\}$.
\end{assum}

Our principal goal is to design a controller to achieve constrained regulation of \eqref{eq:sys} to the origin.
To this end, we propose a tube-based approach using stabilizing terminal conditions based on finite-step contractive sets.

% ------------------------------------------------------------------------------

\subsection{Finite-step contraction}

\begin{defn}\label{def:gs_fstep_contr}
    Let $M\geq 1$ be an integer, let $\lambda\in[0,1)$, let $\mathbf{S}_M=\left\{S_0,\dots,S_{M-1}\right\}$ be a sequence of PC-sets, and define $\sigma(k):=k\bmod M$.
    The PC-set $S_0\subseteq \mathbb{X}$ is called controlled $(M,\lambda)$-contractive, if there exists
    a periodic control law $u(k) = \kappa_{\sigma(k)}\left(x(k),\theta(k)\right)$ with
    $\kappa_i:S_i\times \Theta \rightarrow \mathbb{U}$, $i\in\mathbb{N}_{[0,M-1]}$ such that:
    \begin{subequations}\label{eq:gs_fstep_contr}
        \begin{align}
            &\text{\textbullet}\ \forall i\in\mathbb{N}_{[0,M-2]},\forall x\in S_i,\forall \theta\in\Theta:\\\nonumber
                &\hspace{2em} A(\theta)x+B\kappa_i(x,\theta)\in S_{i+1}\\
            &\text{\textbullet}\ \forall x\in S_{M-1},\forall \theta\in\Theta:\label{eq:gs_fstep_contr:b}\\\nonumber
                &\hspace{2em} A(\theta)x+B\kappa_{M-1}(x,\theta)\in \lambda S_{0},\\
            &\text{\textbullet}\ \forall i \in\mathbb{N}_{[0,M-1]}:\ \left\{0\right\} \subset S_i \subseteq \mathbb{X}.
        \end{align}
    \end{subequations}
    It is assumed that $\kappa_{\sigma(\cdot)}(\cdot,\cdot)$ is (i) continuous and (ii) positively homogeneous, i.e.,
    $\forall (i,x,\theta,\alpha)\in \mathbb{N}_{[0,M-1]}\times\mathbb{R}^{n_{\mathrm{x}}}\times\Theta\times\mathbb{R}_+:\ \kappa_{i}(\alpha x, \theta) = \alpha \kappa_{i}(x, \theta)$.
\end{defn}

Observe that \eqref{eq:gs_fstep_contr:b} means that contraction of $S_0$ is achieved after $M$ time instances.
Sets satisfying these properties were first used in reference governor design \cite{Gilbert2002}, and
can also be interpreted as an instance of the positively invariant families from \cite{Artstein2011}.
Finite-step invariant sets were used in (nominal) non-linear MPC in \cite{Lazar2015},
and the framework of \cite{Gondhalekar2011} employs periodically invariant sets in MPC of linear periodic systems.
Finite-step invariant \emph{ellipsoids} for LPV systems were introduced in \cite{Lee2005}, and applied in a stabilizing open-loop min-max algorithm. 
A practical controller based on these ellipsoids was given in \cite{Lee2006}, where only a single free control action is being optimized on-line.
In the current paper, sets satisfying Definition~\ref{def:gs_fstep_contr} will be employed in the construction of a stabilizing tube-based MPC for LPV systems.

If $S_0$ is a polytope the periodic control laws in Definition~\ref{def:gs_fstep_contr} can always be selected as gain-scheduled vertex controllers,
because -- by convexity -- existence of suitable controls on the vertices of $S_i\times\Theta$ implies existence of suitable controls over the full sets $S_i\times\Theta$, $i\in\mathbb{N}_{[0,M-1]}$
(compare, e.g., \cite[Corollaries~4.46~and~7.8]{Blanchini2008}).
Finally, the closed-loop set-valued dynamics of \eqref{eq:sys} under a given local periodic controller $\kappa_{\sigma(\cdot)}(\cdot,\cdot)$ are
\begin{multline}\label{eq:set_cl_dyn}
    X(k+1) = G\left(k, X(k)|\kappa\right) \\= \left\{A(\theta)x+B\kappa_{\sigma(k)}(x,\theta)\mid x\in X(k),\ \theta\in\Theta\right\}.
\end{multline}
The local uncertain closed-loop dynamics \eqref{eq:set_cl_dyn} are fundamentally different from those in nominally stabilizing MPC \cite{Mayne2000}.
Hence, constructing a suitable terminal cost is a challenge which will be addressed in this paper.

% ==============================================================================

\section{TMPC with finite-step stabilizing conditions}\label{sec:tmpc}

The algorithm constructs, at each time instant $k\in\mathbb{N}$, a so-called constraint invariant tube \cite{Brunner2013,Hanema2016:cdc:final}.

\begin{defn}\label{def:tube}
    A constraint invariant tube for the constraint set $(\mathbb{X},\mathbb{U})\subset\mathbb{R}^{n_{\mathrm{x}}}\times\mathbb{R}^{n_{\mathrm{u}}}$ is defined as
    \[\mathbf{T}_k := \left(\left\{X_{0|k}, \dots,X_{N|k}\right\},\left\{\Pi_{0|k},\dots,\Pi_{N-1|k}\right\}\right)\]
    where $X_{i|k}\subset\mathbb{R}^{n_{\mathrm{x}}}, i\in\mathbb{N}_{[0,N]}$ are sets and $\Pi_{i|k}: X_{i|k}\times \Theta_{i|k} \rightarrow\mathbb{U}, i\in\mathbb{N}_{[0,N-1]}$ are control laws
    satisfying the condition $\forall (x,\theta) \in X_{i|k} \times \Theta_{i|k}: A(\theta)x+B \Pi_{i|k}(x,\theta)\in X_{i+1|k}\cap \mathbb{X}$.
    The sequence of sets $\mathbf{X}_k$ is called the state tube, and each set $X_{i|k}$ is called a cross section.
\end{defn}

Since $\theta(k)$ can be measured according to Assumption~\ref{ass:system_assumptions}, normally we have $\forall k\in\mathbb{N}:\Theta_{0|k}=\{\theta(k)\}$.
The rest of the sequence $\mathbf{\Theta}_{k}:=\left\{\Theta_{0|k},\dots,\Theta_{N-1|k}\right\}$ must satisfy the following assumption and could, e.g., describe a known and bounded rate-of-variation on $\theta$ or an ``anticipated'' approximate future scheduling trajectory, see \cite{Hanema2016:cdc:final}.

\begin{assum}\label{ass:anticip}
    (i) At any two successive time instants, the sequences $\mathbf{\Theta}_{k+1}$ and $\mathbf{\Theta}_{k}$ are related such that $\forall i\in\mathbb{N}_{[0,N-2]}:\ \Theta_{i|k+1}\subseteq \Theta_{i+1|k}$ (continuity).
    (ii) It holds $\forall (k,i) \in \mathbb{N}\times \mathbb{N}_{[0,N-1]}:\ \Theta_{i|k}\subseteq \Theta$ (well-posedness).
    (iii) All sets $\Theta_{i|k}$ are polytopes with $q$ vertices, i.e., $\Theta_{i|k}=\mathrm{convh}\{\bar{\theta}^j_{i|k} \mid j \in\mathbb{N}_{[1,q]}\}$.
\end{assum}

The above Assumptions \ref{ass:anticip}.(i) and \ref{ass:anticip}.(ii) are critical in obtaining recursive feasibility of the MPC scheme.
Assumption \ref{ass:anticip}.(iii) is invoked merely to simplify notation.
To synthesize tubes satisfying Definition~\ref{def:tube} on-line, the cross sections and control laws must be finitely parameterized.
We introduce the 2-tuple of tube parameters
\begin{equation*}
    p_{i|k} = \left(p_{i|k}^X, p_{i|k}^\Pi\right) \in \mathbb{P} = \mathbb{P}_X\times\mathbb{P}_\Pi \subseteq \mathbb{R}^{q_p^X}\times\mathbb{R}^{q_p^\Pi}
\end{equation*}
where $\left(q_p^X,q_p^\Pi\right)\in\mathbb{N}^2$.
Each parameter $p^X_{i|k}$ uniquely characterizes a cross section $X_{i|k}$ and each $p_{i|k}^\Pi$ defines the corresponding controller $\Pi_{i|k}$.
The set $\mathbb{P}= \mathbb{P}_X\times\mathbb{P}_\Pi$ is called the parameterization class.
In the sequel, it has to be understood that any pair $\left(X_{i|k}, \Pi_{i|k}\right)$ is assumed to be parameterized by a corresponding tube parameter $p_{i|k}\in\mathbb{P}$.
That is, we can always construct a time-dependent function $\bar{P}(\cdot,\cdot)$, mapping tube parameters into corresponding sets and controllers, such that $\left(X_{i|k},\Pi_{i|k}\right)=\bar{P}(k+i, p_{i|k})$.
A suitable parameterization, which will be covered fully in Section~\ref{sec:lpv_construction}, is a periodically time-varying homothetic tube where
$X_{i|k}=z_{i|k}\oplus \alpha_{i|k}S_{\sigma(k+i)}$.
Then, $p_{i|k}^X=(\alpha_{i|k},z_{i|k})$ and $S_i$, $i\in\mathbb{N}_{[0,M-1]}$ are sets chosen off-line.
The control laws $\Pi_{i|k}$ are parameterized as the vertex controllers induced by the sets $X_{i|k}$, so that each $p_{i|k}^\Pi$ corresponds to a finite number of control actions.
The tube construction can be formulated as the following optimization problem, to be solved on-line:
\begin{equation}\label{eq:tube_synth}
    \newcommand{\nhs}{\hspace{-5em}}
    \begin{aligned}
    &V\left(k, x_{0|k}, \mathbf{\Theta}_k\right) =\\
        &   \underset{\mathbf{d}_k\in\mathbb{D}}{\text{min}}
        &&  \nhs \sum_{i=0}^{N-1} \ell(X_{i|k}, \Pi_{i|k}) + F_{k}\left(X_{N|k}\right)\\
        &   \ \text{s.t.}
        &&  \nhs \forall i\in \mathbb{N}_{[0,N-1]}:\; \forall x\in X_{i|k},\; \forall \theta \in \Theta_{i|k}:\\
        &&& \nhs\hspace{2.5ex}  A(\theta)x + B\Pi_{i|k}(x,\theta) \in X_{i+1|k}\cap \mathbb{X},\\
        &&& \nhs X_{0|k} = \{x_{0|k}\} ,\ X_{N|k} \subseteq X_{f|k}\subseteq \mathbb{X},
    \end{aligned}
\end{equation}
where $\ell(\cdot,\cdot)$ is the stage cost chosen to meet some desired objective,
and where the time-varying terminal set $X_{f|k}$ and terminal cost $F_k(\cdot)$ are selected to guarantee feasibility and stability.
The decision variable consists of the sequence of tube parameters and is therefore a tuple $\mathbf{d}_{k} = \left(p_{0|k}^X,p_{0|k}^\Pi,\dots,p_{N-1|k}^X, p_{N-1|k}^\Pi, p_{N|k}^X\right)\in\mathbb{D}=\mathbb{P}_X^{N+1}\times\mathbb{P}_\Pi^N$.
In \eqref{eq:tube_synth}, the sets $X_{i|k}$ and controllers $\Pi_{i|k}$ are functions of these tube parameters, but this dependency is omitted from the notation for brevity.
The state measurement at time $k$ is captured in the constraint $X_{0|k} = \{x_{0|k}\}$.
Because the value $\theta(k)$ is measured exactly, the first control law always reduces to a single control action, i.e., $\Pi_{0|k}(x,\theta)=u_{0|k}$.
After solving \eqref{eq:tube_synth}, we set $u(k)=u_{0|k}$ and repeat the optimization at the next sample.
In the sequel, we use a worst-case homogeneous stage cost
\begin{equation}\label{eq:stage_cost}
    \begin{aligned}
        &\ell\left(X_{i|k}, \Pi_{i|k}\right)\\
        &\quad = \max_{(x,\theta)\in X_{i|k}\times \Theta_{i|k}} \left(\|Q x\| + \|R \Pi_{i|k}(x,\theta)\|\right) \\
    \end{aligned}
\end{equation}
where $Q\in\mathbb{R}^{n_{\mathrm{x}}\times n_{\mathrm{x}}}$ and $R\in\mathbb{R}^{n_{\mathrm{u}}\times n_{\mathrm{u}}}$ are tuning parameters.
Let a sequence  $\mathbf{S}_M$ of polytopic controlled $(M,\lambda)$-contractive sets satisfying Definition~\ref{def:gs_fstep_contr} be given.
Then, we select the periodically time-varying terminal set as
\begin{equation}\label{eq:termset}
    X_{f|k} = S_{\sigma(k+N)}.
\end{equation}
To guarantee recursive feasibility the following assumption, an extended variant of \cite[Assumption~7]{Brunner2013}, on the tube parameterization is necessary.
\begin{assum}\label{ass:para}
The terminal set and the associated local controller are ``homogeneously parameterizable'' in $\mathbb{P}$, i.e., 
$\forall (k,\gamma) \in\mathbb{N}\times\mathbb{R}_+:\ \exists p_{f|k}\in\mathbb{P}$ such that $\bar{P}\left(k, p_{f|k}\right) = \gamma \left(S_{\sigma(k)},\kappa_{\sigma(k)}\right)$.
\end{assum}

Later, in Section~\ref{sec:lpv_construction}, a concrete parameterization is given which satisfies Assumption~\ref{ass:para}.
Now, recursive feasibility of \eqref{eq:tube_synth} can be shown.

\begin{prop}\label{prop:feas}
    Let $\mathbf{S}_M$ be a sequence of controlled $(M,\lambda)$-contractive sets for \eqref{eq:sys} according to Definition~\ref{def:gs_fstep_contr}, and
    let the associated closed-loop dynamics $G(\cdot,\cdot|\kappa)$ be as in \eqref{eq:set_cl_dyn}.
    Define the terminal set $X_{f|k}$ as in \eqref{eq:termset}.
    Suppose that Assumptions~\ref{ass:anticip} and~\ref{ass:para} are satisfied.
    Then the TMPC defined by \eqref{eq:tube_synth} is recursively feasible.
\end{prop}

To guarantee stability of the MPC scheme, an appropriate terminal cost has to be constructed.
The first step is to find a Lyapunov-type function which is monotonically decreasing along the set-valued trajectories of \eqref{eq:set_cl_dyn}.
For this, we need the following finite-step decrease property of the function $\Psi_{S_i}(\cdot)$.
The abbreviated notations $\psi_i(\cdot) := \psi_{S_i}$ and $\Psi_i(\cdot) := \Psi_{S_i}(\cdot)$ are used in the sequel.

\begin{lem}\label{lem:gauge_fstep}
    Let $\mathbf{S}_M$ be a sequence of controlled $(M,\lambda)$-contractive sets for \eqref{eq:sys} in the sense of Definition~\ref{def:gs_fstep_contr}.
    Define the resulting closed-loop dynamics $G(\cdot,\cdot|\kappa)$ as in \eqref{eq:set_cl_dyn}.
    Then, $\Psi_{\sigma(k)}\left(\cdot\right)$ satisfies $\forall k\in\mathbb{N}$ and $\forall X\subseteq S_{\sigma(k)}$:
    \begin{equation*}
        \Psi_{\sigma(k+1)}\left(G(k, X|\kappa)\right) \leq
        \begin{cases}
            \Psi_{\sigma(k)}\left(X\right), & \hspace{-0.5em}\sigma(k)\in\mathbb{N}_{[0,M-2]},\\
            \lambda \Psi_{\sigma(k)}\left(X\right), & \hspace{-0.5em}\sigma(k)=M-1.
        \end{cases}
    \end{equation*}
\end{lem}

The above lemma can be exploited to construct a Lyapunov-type function enabling the computation of a stabilizing terminal cost for \eqref{eq:tube_synth}.
The following proposition applies a suitably modified version of the construction of \cite[Theorem~20]{Geiselhart2014a} to sequences of sets, yielding the desired function.

\begin{prop}\label{prop:periodic_lyap}
    Suppose that the conditions from Lemma~\ref{lem:gauge_fstep} are satisfied.
    Then, the function \[W\left(k,X\right) := \left(M + \left(\lambda-1\right)\sigma(k)\right)\Psi_{\sigma(k)}\left(X\right)\] is a Lyapunov-type function for the dynamics (\ref{eq:set_cl_dyn}), i.e., it satisfies the following properties:
    \begin{enumerate}[(i)]
        \item $\exists s_6,s_7\in\mathcal{K}_\infty$ such that $\forall k\in\mathbb{N}:\ \forall X\in\mathcal{C}^n:\ s_6\left(d_H^0\left(X\right)\right) \leq W\left(k,X\right) \leq s_7\left(d_H^0\left(X\right)\right)$ holds,
        
        \item $\exists \varrho(k):\mathbb{N}\rightarrow[0,1)$ such that
        $\forall k\in\mathbb{N}:\ \forall X\subseteq S_{\sigma(k)}:\ W\left(k+1,G\left(k,X|\kappa\right)\right) \leq \varrho(k) W\left(k, X\right)$,
        
        \item $\exists \varrho\in[0,1)$ such that $\forall k\in\mathbb{N}:\ \forall X\subseteq S_{\sigma(k)}:\ W\left(k+1,G\left(k,X|\kappa\right)\right) \leq \varrho W\left(k, X\right)$.
    \end{enumerate}
\end{prop}

The next step towards a stability proof is to construct a scaling of $W(\cdot,\cdot)$ to obtain a terminal cost for \eqref{eq:tube_synth}.
For all $i\in\mathbb{N}_{[0,M-1]}$, let
\begin{equation}\label{eq:stage_ubound}
    \begin{aligned}
        \bar{\ell}_i &= \underset{(x,u)\in S_i\times\mathbb{U}}{\text{max}}
        \left(\|Q x\| + \|Ru\|\right)\ \text{s.t.}\ \forall \theta\in\Theta: \\
        & \phantom{=}  \begin{cases}
            A(\theta)x + B u\in S_{i+1}, & i\in\mathbb{N}_{[0,M-2]},\\
            A(\theta)x + B u\in \lambda S_{0}, & i=M-1.
        \end{cases}
    \end{aligned}
\end{equation}
From Proposition~\ref{prop:periodic_lyap} the next Corollary follows directly.

\begin{cor}\label{cor:vf}
    Let $\bar{\ell}_i$ be as in \eqref{eq:stage_ubound} and define $\bar{\ell}=\max_{i\in\mathbb{N}_{[0,M-1]}} \bar{\ell}_i$.
    Define $W(k,X)$ and $\varrho$ as in Proposition~\ref{prop:periodic_lyap} and $X_{f|k}$ as in \eqref{eq:termset}.
    Then the function
    \begin{equation}\label{eq:vf}
        \overline{W}\left(k, X\right) := \frac{\bar{\ell}}{1-\varrho} W\left(k, X\right)
    \end{equation}
    satisfies $\forall k\in\mathbb{N}$ and $\forall X\subseteq S_{\sigma(k)}$:
    \begin{equation*}
        \overline{W}\left(k+1, G\left(k,X|\kappa\right)\right) - \overline{W}\left(k, X\right) \leq -\bar{\ell} W\left(k, X\right).
    \end{equation*}
    Furthermore, $\forall k\in\mathbb{N}: 1 \leq W\left(k+N, X_{f|k}\right) \leq M$.
\end{cor}

Before proving asymptotic stability of the TMPC scheme, the following assumptions on the stage cost and value function are required.

\begin{assum}\label{ass:stage_kinf}
    (i) Let $(k,p)\in\mathbb{N}\times\mathbb{P}$ such that $\bar{P}\left(k,p\right)=\left(X_k,\Pi_k\right)$ with $\Pi_k:X\times\Theta\rightarrow\mathbb{U}$.
    Then there exists a $\mathcal{K}_\infty$-function $s_8$ such that
    $s_8\left(d_H^0\left(X_k\right)\right)\leq \ell(X_k, \Pi_k)$.
    (ii) There exist $\mathcal{K}_\infty$-functions $s_{9},s_{10}$ such that for all $k\in\mathbb{N}$ and for all $x_{0|k}\in\mathbb{R}^{n_\mathrm{x}}$ for which \eqref{eq:tube_synth} is feasible it holds $s_{9}\left(\|x_{0|k}\|\right) \leq V\left(k,x_{0|k},\mathbf{\Theta}_k\right) \leq s_{10}\left(\|x_{0|k}\|\right)$.
\end{assum}

In Section~\ref{sec:lpv_construction} it is proven that, for a certain choice of tube parameterization, the stage cost \eqref{eq:stage_cost} and the value function of \eqref{eq:tube_synth} indeed satisfy the above assumptions.
Now we state the main result.

\begin{thm}\label{thm:stab}
    Suppose that the conditions of Proposition~\ref{prop:feas} and Assumption~\ref{ass:stage_kinf} are satisfied.
    Let $F_k(\cdot):=\overline{W}\left(k+N,\cdot\right)$ according to \eqref{eq:vf}.
    Then the TMPC defined by \eqref{eq:tube_synth} asymptotically stabilizes the origin.
\end{thm}

% ==============================================================================-

\section{Implementation details}\label{sec:lpv_construction}

In this section, it is shown how the general results presented previously can be used in practice by developing a specific parameterization that satisfies Assumptions~\ref{ass:para} and \ref{ass:stage_kinf}.
To satisfy Assumption~\ref{ass:para}, consider a ``periodic'' variant on the homothetic parameterization \cite{Langson2004,Brunner2013,Rakovic2012a} by parameterizing the tube cross sections as
\begin{equation}\label{eq:para_lpv_tube_sec}
    X_{i|k} = z_{i|k} \oplus \alpha_{i|k} S_{\sigma(k+i)}
\end{equation}
where $z_{i|k}\in\mathbb{R}^{n_{\mathrm{x}}}$ and $\alpha_{i|k}\in\mathbb{R}_+$ are optimized on-line.
Thus, each cross section $X_{i|k}$ is considered homothetic to $S_{\sigma(k+i)}$ with center $z_{i|k}$ and scaling $\alpha_{i|k}$.
The sets $S_i$, $i\in\mathbb{N}_{[0,M-1]}$ are the same as in (\ref{eq:termset}) and they are polytopes represented by the convex hull of $t_i$ vertices as
\begin{equation}\label{eq:s_convh}
    \forall i\in\mathbb{N}_{[0,M-1]}: S_i = \mathrm{convh}\left\{\bar{s}_i^1,\dots,\bar{s}_i^{t_i}\right\}.
\end{equation}
The associated control laws are parameterized as gain-scheduled vertex controllers, i.e.,
\begin{equation}\label{eq:gs_vertex_control}
    \Pi_{i|k}(x,\theta) = \sum_{j=1}^{t_{\sigma(k+i)}} \zeta_j \sum_{l=1}^q \eta_l u^{(j,l)}_{i|k}
\end{equation}
where $u_{i|k}^{(j,l)}\in\mathbb{U}$ are control actions and $\zeta\in\mathbb{R}^{t_{\sigma(k+i)}}$ and $\eta\in\mathbb{R}^{q}$ are convex multipliers in the state- and scheduling spaces, respectively.
At each prediction time instant $k+i$, the control $u_{i|k}^{(j,l)}$ is associated with the \mbox{$j$-th} vertex of the cross section $X_{i|k}$ and the \mbox{$l$-th} vertex of the relevant scheduling set (see Assumption~\ref{ass:anticip}).
The tube parameters $p_{i|k}=\bigl(p_{i|k}^X,p_{i|k}^\Pi\bigr)$ corresponding to the given parameterization are
\begin{equation*}
    p_{i|k}^X = \left(\alpha_{i|k}, z_{i|k}\right),\ 
    p_{i|k}^\Pi = \left(u_{i|k}^{(1,1)},\dots,u_{i|k}^{(t_{\sigma(k+i)},q)}\right).
\end{equation*}
Because the representation \eqref{eq:sys} has a constant $B$-matrix, it is sufficient to verify the existence of the individual control actions $u_{i|k}^{(j,l)}$ to establish the existence of a tube satisfying Definition~\ref{def:tube}, i.e., computation of the convex multipliers $(\zeta,\eta)$ is not necessary.
With this parameterization, the stage cost \eqref{eq:stage_cost} becomes
\begin{equation}\label{eq:stage_cost_homothetic}
    \begin{aligned}
        &\ell(X_{i|k}, \Pi_{i|k})\\
                &\quad = \max_{j\in\mathbb{N}_{[1,t_{\sigma(k+i)}]}, l\in\mathbb{N}_{[1,q]}} \left(\|Q \bar{x}_{i|k}^j\| + \|R u_{i|k}^{(j,l)}\|\right)
    \end{aligned}
\end{equation}
where $\bar{x}_{i|k}^j = z_{i|k} + \alpha_{i|k}\bar{s}_{\sigma(k+i)}^j$.
The scaling $\bar{\ell}$ in Corollary~\ref{cor:vf} can be efficiently computed as follows.
For all $(i,l)\in \mathbb{N}_{[0,M-1]} \times \mathbb{N}_{[1,q]}$ and all corresponding $j \in\mathbb{N}_{[1,t_i]}$, compute the control actions
\begin{align*}
    u_f^{(i,j,l)} &= \underset{u\in \mathbb{U}}{\text{arg\ min}}
    \left(\|Q\bar{s}_i^j\| + \|Ru\|\right)\\
    &\phantom{=} \text{s.t.}\ \begin{cases}
        A(\bar{\theta}^l)\bar{s}_i^j + Bu\in S_{i+1}, & i\in\mathbb{N}_{[0,M-2]},\\
        A(\bar{\theta}^l)\bar{s}_i^j + Bu\in \lambda S_{0}, & i=M-1,
    \end{cases}
\end{align*}
to obtain a local periodic vertex control law which is feasible and asymptotically stabilizing on $\mathbf{S}_M$.
Then, the constants $\bar{\ell}_i$ are directly found by computing
\begin{multline}\label{eq:stage_ubound_lpv}
    \forall i\in\mathbb{N}_{[0,M-1]}:\\ \bar{\ell}_i = \max_{j\in\mathbb{N}_{[1,t_i]}, l\in\mathbb{N}_{[1,q]}} \left(\|Q\bar{s}^j_i\| + \|Ru_f^{(i,j,l)}\|\right),
\end{multline}
and the next Lemma follows.
\begin{lem}\label{lem:cost}
    Suppose that the tuning parameter $Q\in\mathbb{R}^{n_\mathrm{x}\times n_\mathrm{x}}$ is strictly positive definite and therefore of rank $n_\mathrm{x}$.
    Then, the stage cost \eqref{eq:stage_cost_homothetic} and value function of \eqref{eq:tube_synth} satisfy Assumption~\ref{ass:stage_kinf}.
\end{lem}   

It has now been shown that the parameterization defined in this section satisfies all the necessary assumptions from Section~\ref{sec:tmpc}.
The next conclusion follows directly.

\begin{cor}\label{cor:homothetic_stab}
    The LPV TMPC algorithm with tube parameterization \eqref{eq:para_lpv_tube_sec}-\eqref{eq:gs_vertex_control} is recursively feasible and asymptotically stabilizing.
\end{cor}

With the choice of stage cost \eqref{eq:stage_cost} and under the assumption that all involved sets are polytopes, the optimization problem \eqref{eq:tube_synth} is a \emph{linear program} (LP).
Its complexity, in terms of the number of decision variables and constraints, scales linearly in the prediction horizon $N$.
As they are polytopes, each set $S_i$ in \eqref{eq:s_convh} can be equivalently represented in a half-space form with $r_i$ hyperplanes.
Half-space representations of $\mathbb{X}$ and $\mathbb{U}$ are also assumed to be available with $r_\mathbb{X}$ and $r_\mathbb{U}$ hyperplanes, respectively.
According to the discussion below Definition~\ref{def:tube} and to Assumption~\ref{ass:anticip}, let $q(i)=1$ when $i=0$ and $q(i)=q$ otherwise.
Then, by using the half-space representations to verify set inclusions similarly to the implementation described in \cite{Hanema2016:cdc:final},
an LP can be formulated which has
\begin{equation*}
    n_\mathrm{d}(k, N) = 1 + \left(N + 1\right)\left(n_\mathrm{x} + 3\right) + \sum_{i=0}^N n_\mathrm{u} q(i) t_{\sigma\left(k+i\right)}
\end{equation*}
decision variables, and
\begin{multline*}
    n_\mathrm{ineq}(k, N) = 1 + r_{\sigma(k+N)}t_{\sigma(k+N)}
    + \sum_{i=0}^N \Bigl(\bigl(r_\mathbb{X}q(i) + r_\mathbb{U}q(i) \\+ r_{\sigma\left(k+i+1\right)}q(i) + 2n_\mathrm{x} + 2n_\mathrm{u}q(i)\bigr)t_{\sigma(k+i)}\Bigr)
\end{multline*}
linear inequality constraints.
The initial state constraint, finally, gives rise to $n_\mathrm{x}+1$ linear equality constraints.
%An illustration of the resulting problem size on the numerical example of Section~\ref{sec:example} is shown in Table~\ref{tab:sim_computations}.
Note that alternative formulations of the LP avoiding the computation of the hyperplane representations of $S_i$, $i\in\mathbb{N}_{[0,M-1]}$, can be constructed, however their exact complexities were not studied in the context of this work.

The construction of the sequence of finite-step contractive sets $\mathbf{S}_M$ for an LPV system can be done in several ways.
One can pick an arbitrary PC-set $S_0$ and find the smallest $M$ for which a sequence $\mathbf{S}_M$ exists, using a straightforward extension of the algorithm for the LTI case from \cite{Athanasopoulos2013}.
Due to exponential complexity in $M$, this method is only practical when contraction can be achieved for small $M$.
Alternatively, it is possible to first determine any stabilizing controller for \eqref{eq:sys}.
Then again we can choose an arbitrary PC-set $S_0$ and propagate this set forwards under the resulting closed-loop dynamics until finite-step contraction is achieved, as proposed in \cite{Lazar2015}.
The number of vertices of the sets in the resulting sequence $\mathbf{S}_M$ grows exponentially in principle, but often many vertices are redundant and can be eliminated using standard algorithms: a similar technique was employed in \cite{Athanasopoulos2014} for the stability analysis of switched systems.

% ==============================================================================

\section{Numerical example}\label{sec:example}

The approach is now demonstrated on an example.
We consider a second-order LPV system defined in the state-space form of \eqref{eq:sys} with two scheduling variables where
\begin{align*}
    &A_0 = \begin{bmatrix}1 & 1\\0 & 1\end{bmatrix},\ &&A_1 = \begin{bmatrix}0.08 & -0.6\\0.4 & 0.1\end{bmatrix},\\
    &A_2 = \begin{bmatrix}0.23 & 0\\0 & -0.32\end{bmatrix},\ &&B =\;\, \begin{bmatrix}0\\1\end{bmatrix}
\end{align*}
and furthermore
\begin{align*}
    \Theta &= \left\{\theta\in\mathbb{R}^2 \mid \|\theta\|\leq 1\right\},\
    \mathbb{U} = \left\{u\in\mathbb{R}\mid |u|\leq 6\right\},\\
    \mathbb{X} &= \left\{x\in\mathbb{R}^2\mid |x_1|\leq 4, |x_2|\leq 10\right\}.
\end{align*}
The MPC tuning parameters are $N=8$, $Q=I$, and $R=0.25$.
This tuning assigns a low weight to the control input, leading to a fast response.
For simplicity, we set $\Theta_{i|k}=\Theta$ for all $(k,i)$.

\begin{figure}
	\centering
	\includegraphics[width=\columnwidth]{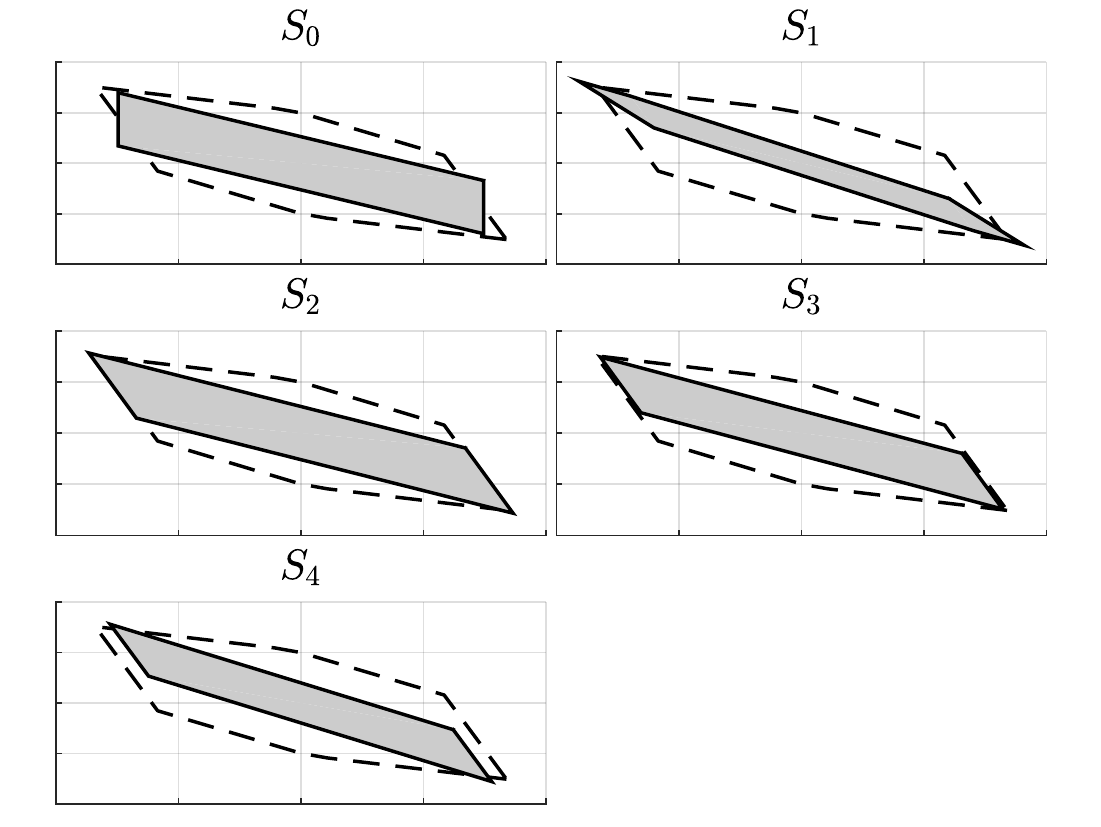}
	\caption{Constructed sequence of $(5,0.95)$-contractive sets (solid) compared with the maximal $0.95$-contractive set (dashed).\label{fig:sim_sm}}
\end{figure}

A set $S_0$ was chosen which leads to a sequence $\mathbf{S}_M$ of $(5,0.95)$-contractive sets, as depicted in Figure~\ref{fig:sim_sm}.
The set $S_0$ was designed with 4 vertices,
and all subsequent sets also have 4 vertices except for $S_1$, which has 6.
For comparison, the maximal controlled $0.95$-contractive set was also calculated using the algorithm from \cite{Blanchini2008} and it has 8 vertices.

\begin{figure}
    \centering
    \includegraphics[width=\columnwidth]{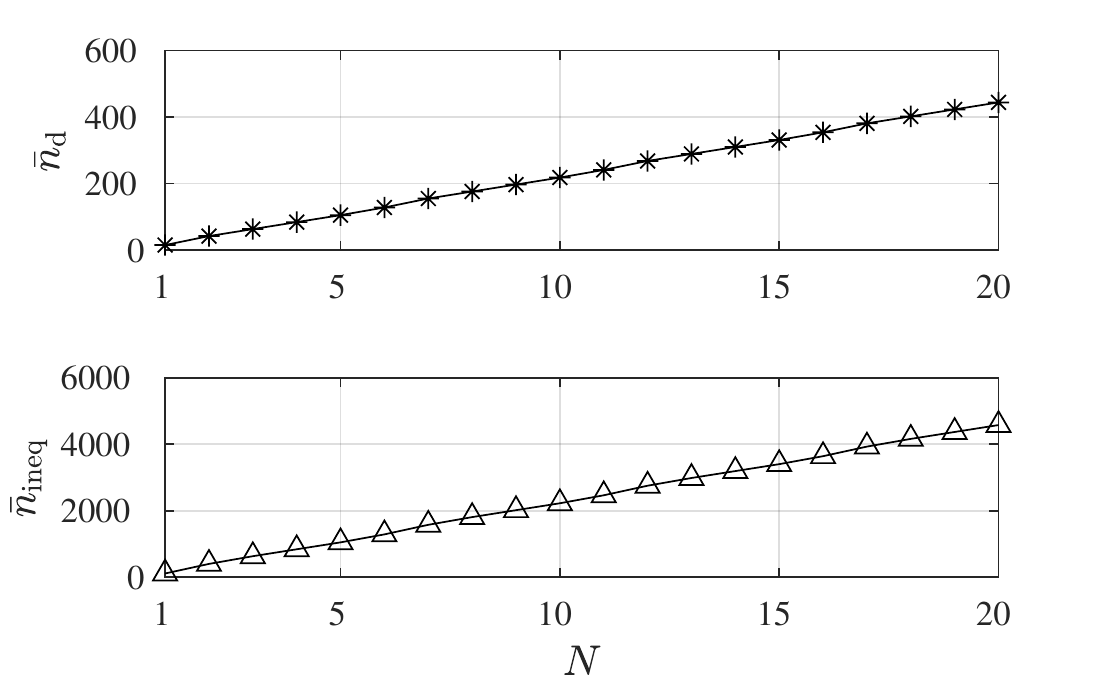}
    \caption{Max. number of variables $\bar{n}_\mathrm{d}(N)=\max_k n_\mathrm{d}(k,N)$ and constraints $\bar{n}_\mathrm{ineq}(N)=\max_k n_\mathrm{ineq}(k,N)$.\label{fig:dims}}
\end{figure}

The relative difference in computational load of the resulting TMPC algorithm, based on an LP implementation where both the vertex- and hyperplane representations of the sets were used, is displayed in Table~\ref{tab:sim_computations}.
The simulations were carried out on a $3.6$ GHz Intel Core i7-4790 with 8 GB RAM, running Arch Linux, and using the Gurobi 7.0.2 LP solver with its default settings.
\begin{table}
	\begin{tabular}{l|cccc}
    	$(M,\lambda)$ & $n_\mathrm{d}$ & $n_\mathrm{ineq}$ & Avg. (max.) time\\\hline
    	$(1,0.95)$ & 276 & 4034 & 14 (20) [ms] &\\
    	$(5,0.95)$ & 168-176 & 1674-1810 & 6 (8) [ms] &\\
	\end{tabular}
	\caption{Illustration of complexity: number of decision variables, number of inequality constraints, and solver time per sample.\label{tab:sim_computations}}
    \hrule
\end{table}
Because the complexity of the terminal set in the $(5,0.95)$-contractive case is time-dependent, the number of variables and constraints varies periodically between the numbers shown.
To illustrate their linear growth, the maximum number of variables and constraints for the $(5,0.95)$-contractive case is calculated as a function of $N$ and shown in Figure~\ref{fig:dims}.

\begin{figure}
	\centering
	\includegraphics[width=\columnwidth]{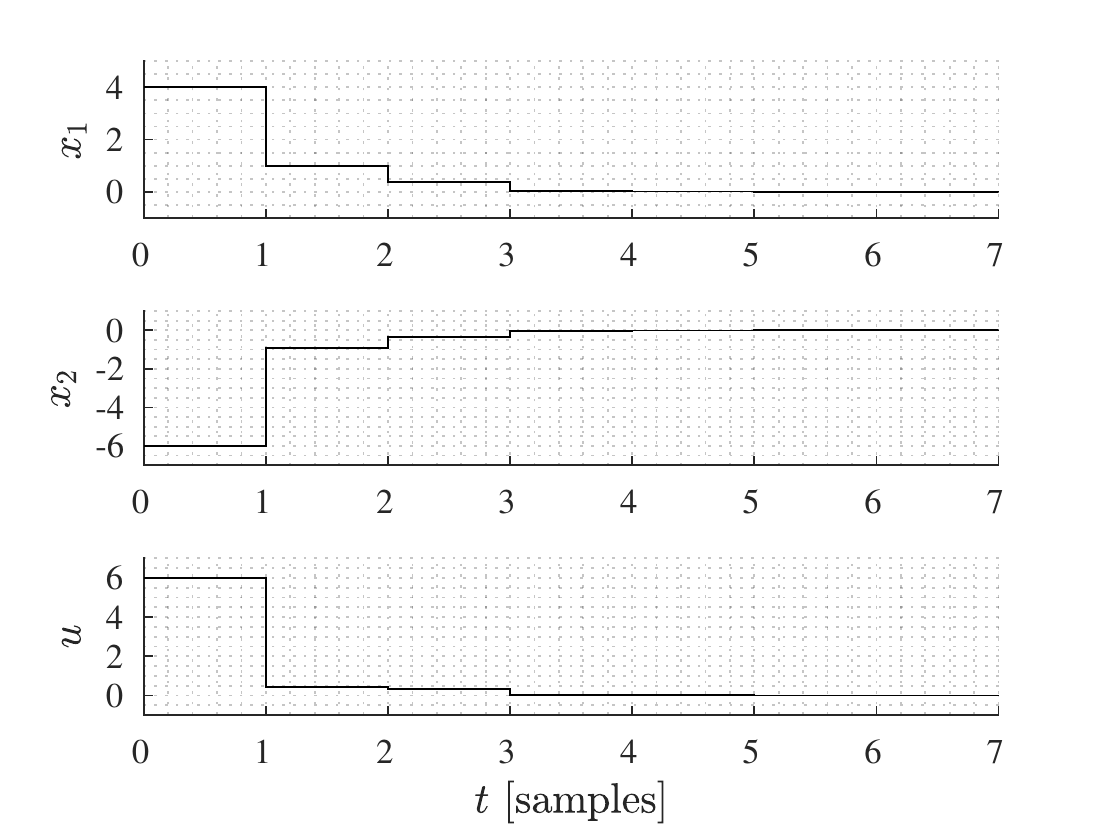}
	\caption{Closed-loop state- and input trajectories with finite-step terminal condition.\label{fig:sim_out}}
\end{figure}

\begin{figure}
	\centering
	\includegraphics[width=\columnwidth]{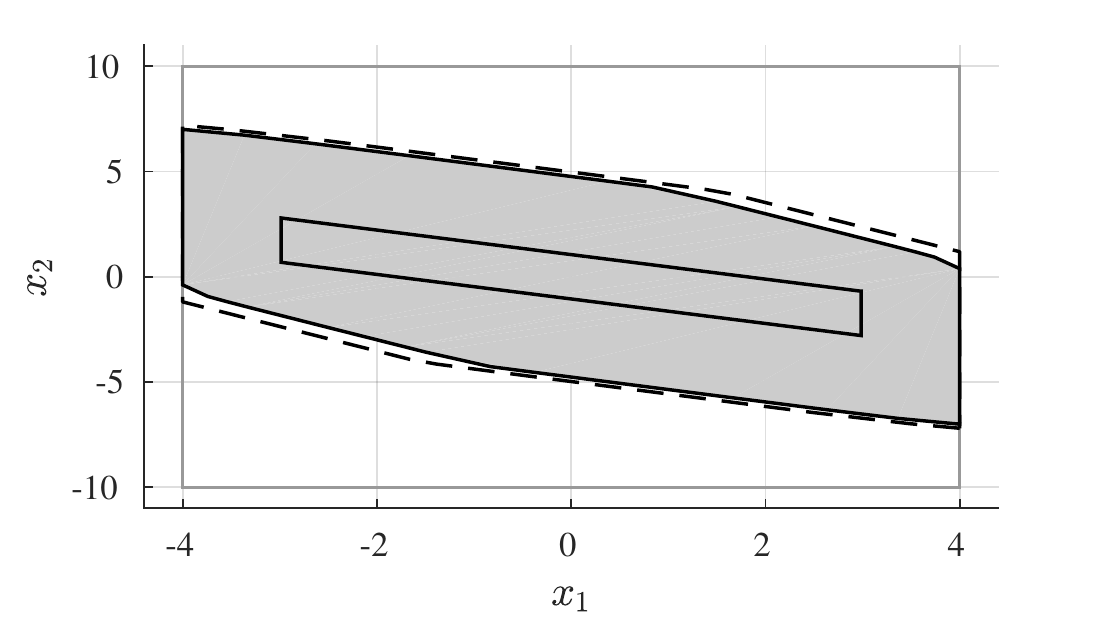}
	\caption{Approximate domains of attraction with finite-step terminal condition (filled) and with maximal contractive terminal set (dashed line),
        for $\theta(0)=\bigl[1\ -1\bigr]^\top$. The innermost set (solid line) is $S_0$ from Figure~\ref{fig:sim_sm}, and the outer box represents the state constraints.\label{fig:sim_feas}}
\end{figure}

An example closed-loop output trajectory of the controller with finite-step terminal condition is shown in Figure~\ref{fig:sim_out}.
The scheduling trajectory was generated randomly and the initial state was $x(0)=\bigl[\begin{array}{cc}4 & -6\end{array}\bigr]$, i.e., taken at the boundary of the state constraint set.
As expected, the system's state variables are steered to the origin and input- and state constraints are satisfied.
We also compare the achieved domains of attraction of the controller with the finite-step terminal condition to that of the controller from \cite{Hanema2016:cdc:final}, which uses the maximal $0.95$-contractive terminal set (Figure~\ref{fig:sim_feas}).
The feasible set was calculated for a fixed initial value $\theta(0)=\bigl[\begin{array}{cc}1 & -1\end{array}\bigr]^\top$.
In the present case, the reduction in computational load due to the lesser complexity of the sets in $\mathbf{S}_M$ is paid for by a marginally smaller feasible set.


% ==============================================================================

\section{Conclusion}

The present work has introduced finite-step terminal conditions in tube-based MPC for LPV systems.
It was shown that, under certain assumptions on the tube parameterization, the method is recursively feasible.
A new Lyapunov-like function on periodic sequences of PC-sets was constructed:
it was subsequently used to derive a terminal cost, enabling a proof of closed-loop asymptotic stability.
Extension to constrained output reference tracking for LPV systems is a future direction of interest.

% ==============================================================================

\section*{Appendix. Proofs}

\paragraph*{Proof of Lemma~\ref{lem:gauge_kinf}.}
Denote $\mathcal{B}_\infty = \left\{x\mid \|x\| \leq 1\right\}$.
Note that $\psi_{\mathcal{B}_\infty}(x)=\|x\|$.
For sets $S_1,S_2\subset \mathbb{R}^n$ with $S_1\subseteq S_2$, it holds $\psi_{S_1}(x)\geq \psi_{S_2}(x)$ for all $x\in\mathbb{R}^n$ \cite[Lemma~1]{Rakovic2012c}.
Because $S$ is a PC set, $\exists a,b\in\mathbb{R}_+$ such that $a\mathcal{B}_\infty \subseteq S \subseteq b\mathcal{B}_\infty$.
Thus, $\forall x\in\mathbb{R}^n:\ b^{-1}\psi_{\mathcal{B}_\infty}(x) \leq \psi_S(x) \leq a^{-1}\psi_{\mathcal{B}_\infty}(x) $, i.e.,
statement (i) holds with $s_1(\xi)=b^{-1} \xi$ and $s_2(\xi)=a^{-1}\xi$.
Next, observe that $\Psi_{\mathcal{B}_\infty}(X)=\sup_{x\in X} \|x\| = d_H^0(X)$.
For sets $S_1,S_2\subset \mathbb{R}^n$ with $S_1\subseteq S_2$, it similarly holds $\Psi_{S_1}(X)\geq \Psi_{S_2}(X)$ for all $X\in\mathcal{C}^n$.
Hence, $\forall X\in\mathcal{C}^n:\ b^{-1}\Psi_{\mathcal{B}_\infty}(X) \leq \Psi_S(X) \leq a^{-1}\Psi_{\mathcal{B}_\infty}(X) $, i.e.,
statement (ii) follows with $s_3(\xi)=b^{-1} \xi$ and $s_4(\xi)=a^{-1} \xi$.
\qed
    
\paragraph*{Proof of Proposition~\ref{prop:feas}.}
Suppose that \eqref{eq:tube_synth} is feasible at time $k$ and let
\begin{equation*}
    \mathbf{T}_k^\star = \left(\left\{X_{0|k},\dots,X_{N|k}\right\}, \left\{\Pi_{0|k},\dots,\Pi_{N-1|k}\right\}\right)
\end{equation*}
be the tube resulting from the optimal solution of \eqref{eq:tube_synth} at time $k$.
By construction, $X_{0|k}=\{x_{0|k}\}$
and $\exists \gamma\in[0,1]: X_{N|k}\subseteq \gamma X_{f|k}$.
Note that $\gamma=1$ would be sufficient here, but keeping it variable simplifies the subsequent stability proof of Theorem~\ref{thm:stab}.
After applying $\Pi_{0|k}$ to the system,
by definition of the terminal set and under Assumption~\ref{ass:anticip} a feasible tube at time $k+1$ can be explicitly given as
\begin{align*}
    &\mathbf{T}_{k+1}^\circ = \bigl(\bigl\{X_{0|k+1}, X_{2|k}, \dots, X_{N-1|k}, \gamma X_{f|k},\\
    &\quad \gamma G\left(k+N, X_{f|k}|\kappa\right)\bigr\},
    \bigl\{\Pi_{1|k},\dots, \Pi_{N-1|k}, \gamma \kappa_{N} \bigr\}\bigr),
\end{align*}
where $X_{0|k+1}=\{x_{0|k+1}\}\subset X_{1|k}$,
which implies feasibility of $\Pi_{0|k+1}=\Pi_{1|k}$.
Since \eqref{eq:tube_synth} only optimizes over finitely parameterized sets and controllers, there must exist parameters
$\left(p_{f|k}, p_{f|k+1}\right)\in\mathbb{P}^2$ such that $\bar{P}\left(k+N, p_{f|k}\right)=\gamma \left(X_{f|k},\kappa_{N}\right)$ and
$\bar{P}\left(k+N+1, p_{f|k+1}\right)=\gamma\left(G\left(k+N, X_{f|k}|\kappa\right), \ast\right)$ where $\ast$ denotes an irrelevant quantity.
This is guaranteed by Assumption~\ref{ass:para}, and therefore it follows that \eqref{eq:tube_synth} is feasible at time $k+1$.
\qed

\paragraph*{Proof of Lemma~\ref{lem:gauge_fstep}.}
Let $\partial S$ denote the boundary of a set $S\subset\mathbb{R}^n$.
By Definition~\ref{def:gs_fstep_contr}, $\forall \bar{x}\in\partial S_{\sigma(k)}$:
\begin{equation*}
    G\left(k, \{\bar{x}\}|\kappa\right)\in
    \begin{cases}
        S_{\sigma(k+1)}, & \sigma(k)\in\mathbb{N}_{[0,M-2]},\\
        \lambda S_{\sigma(k+1)}, & \sigma(k)=M-1.
    \end{cases}
\end{equation*}
Now let $x\in S_{\sigma(k)}$.
By definition of the gauge function it holds $x \in \psi_{\sigma(k)}(x)\partial S_{\sigma(k)}$ \cite{Schneider2013c1}.
Thus, $\exists \bar{x}\in\partial S_{\sigma(k)}: x=\psi_{\sigma(k)}(x)\bar{x}$.
By homogeneity it follows directly that
\begin{multline*}
    G\left(k,\{x\}|\kappa\right)=G\left(k, \{\psi_{\sigma(k)}(x)\bar{x}\}|\kappa\right)\\=\psi_{\sigma(k)}(x)G\left(k, \{\bar{x}\}|\kappa\right)
\end{multline*}
and therefore $\forall x\in S_{\sigma(k)}$:
\begin{equation*}
    G\left(k, \{x\}|\kappa\right)\in
    \begin{cases}
        \psi_{\sigma(k)}(x) S_{\sigma(k+1)}, & \sigma(k)\in\mathbb{N}_{[0,M-2]},\\
        \lambda \psi_{\sigma(k)}(x) S_{\sigma(k+1)}, & \sigma(k)=M-1.
    \end{cases}
\end{equation*}
From the above we get that $\forall X\subseteq S_{\sigma(k)}$:
\begin{multline*}
    G\left(k, X|\kappa\right)\subseteq\\
    \begin{cases}
        \sup_{x\in X} \psi_{\sigma(k)}(x) S_{\sigma(k+1)}, & \sigma(k)\in\mathbb{N}_{[0,M-2]},\\
        \lambda \sup_{x\in X} \psi_{\sigma(k)}(x) S_{\sigma(k+1)}, & \sigma(k)=M-1
    \end{cases}
\end{multline*}
and by applying Definition~\ref{def:set_gauge} the desired property follows.
\qed
    
\paragraph*{Proof of Proposition~\ref{prop:periodic_lyap}.}
Since
$\left(M + (\lambda - 1)\sigma(k)\right)$ is a positive number for all $k\in\mathbb{N}$, it follows from Lemma~\ref{lem:gauge_kinf} that
$\exists s_6^i,s_7^i\in\mathcal{K}_\infty$ for each $i\in\mathbb{N}_{[0,M-1]}$ such that $\forall X\in\mathcal{C}^n:\ s_6^{\sigma(k)}\left(d_H^0(X)\right) \leq W(k,X) \leq s_7^{\sigma(k)}\left(d_H^0(X)\right)$.
As the minimum- and maximum over a finite set of $\mathcal{K}_\infty$-functions is again $\mathcal{K}_\infty$, statement (i) holds with
$s_6(\xi) = \min_{i\in\mathbb{N}_{[0,M-1]}} s_6^i(\xi)$ and $s_7(\xi) = \max_{i\in\mathbb{N}_{[0,M-1]}} s_7^i(\xi)$.
For the proof of (ii), consider first that $k$ is such that $\sigma(k)\in\mathbb{N}_{[0,M-2]}$.
Then by Lemma~\ref{lem:gauge_fstep},
$\Psi_{\sigma(k+1)}\left(G\left(k,X|\kappa\right)\right) \leq \Psi_{\sigma(k)}\left(X\right)$, and therefore
\begin{align*}
    &W\left(k+1,G\left(k,X|\kappa\right)\right)\\
    &\quad =    \left(M + \left(\lambda-1\right)\sigma(k+1)\right)\Psi_{\sigma(k+1)}\left(G\left(k,X|\kappa\right)\right)\\
    &\quad \leq \left(M + \left(\lambda-1\right)\sigma(k+1)\right)\Psi_{\sigma(k)}\left(X\right)\\
    &\quad =    \frac{\left(M + \left(\lambda-1\right)\sigma(k+1)\right)}{\left(M + \left(\lambda-1\right)\sigma(k)\right)} W\left(k, X\right).
\end{align*}
Next, let $k$ be such that $\sigma(k)=M-1$.
Again by Lemma~\ref{lem:gauge_fstep}, $\Psi_{\sigma(k+1)}\left(G\left(k,X|\kappa\right)\right) \leq \lambda \Psi_{\sigma(k)}\left(X\right)$, so
\begin{align*}
    W\left(k+1,G\left(k,X|\kappa\right)\right)
    &=    M \Psi_0\left(G\left(M-1,X|\kappa\right)\right)\\
    &\leq \lambda M \Psi_{M-1}\left(X\right)\\
    &=    \frac{\lambda M}{\lambda\left(M-1\right)+1} W(k,X).
\end{align*}
Hence, statement (ii) is satisfied with
\begin{equation*}
    \varrho(k) = \begin{cases}
        \frac{\left(M + \left(\lambda-1\right)\sigma(k+1)\right)}{\left(M + \left(\lambda-1\right)\sigma(k)\right)}, & \sigma(k)\in\mathbb{N}_{[0,M-2]},\\
        \frac{\lambda M}{\lambda\left(M-1\right)+1}, & \sigma(k)=M-1,\\
    \end{cases}
\end{equation*}
and (iii) follows with $\varrho = \max_{k\in\mathbb{N}} \varrho(k) = \varrho(0)$.
\qed

\paragraph*{Proof of Theorem~\ref{thm:stab}.}
Let $G_{f|k}(\cdot):=G\left(k+N, \cdot|\kappa\right)$ according to \eqref{eq:set_cl_dyn}.
Consider the optimal solution $\mathbf{T}_k^\star$ and the feasible, but not necessarily optimal, solution $\mathbf{T}_{k+1}^\circ$ constructed in the proof of Proposition~\ref{prop:feas}.
Further, let $\mathbf{\Theta}_k$ and $\mathbf{\Theta}_{k+1}$ be two anticipated scheduling sequences satisfying Assumption~\ref{ass:anticip}.
By definition of $F_k(\cdot)$, it follows that we can take $\gamma=\Psi_{\sigma(k+N)}\left(X_{N|k}\right)$.
Substitute the solutions $\mathbf{T}_k^\star$ and $\mathbf{T}_{k+1}^\circ$ in the cost function of \eqref{eq:tube_synth} and compute the difference between the value functions at time $k$ and time $k+1$ to obtain
\begin{align*}
    \Delta V_k &= V\left(k+1,x_{0|k+1},\mathbf{\Theta}_{k+1}\right) - V\left(k,x_{0|k},\mathbf{\Theta}_k\right)\\
    &\leq \ell\left(X_{0|k+1}, \Pi_{1|k}\right) + \gamma \ell\left(X_{f|k}, \Pi_{f|k}\right)\\
    &\phantom{\leq} + \gamma F_{k+1}\left(G_{f|k}\left(X_{f|k}\right)\right) - F_{k}\left(X_{N|k}\right)\\
    &\phantom{\leq} + \sum_{i=2}^{N-1} \ell\left(X_{i|k}, \Pi_{i|k}\right) - \sum_{i=0}^{N-1} \ell\left(X_{i|k}, \Pi_{i|k}\right).
\end{align*}
Observe that $X_{0|k+1}=\{x_{0|k+1}\}\subset X_{1|k}$, so\\ $\ell\left(X_{0|k+1},\Pi_{1|k}\right)\leq \ell\left(X_{1|k},\Pi_{1|k}\right)$ and therefore
\begin{align*}
    \Delta V_k &\leq \sum_{i=1}^{N-1} \ell\left(X_{i|k}, \Pi_{i|k}\right) - \sum_{i=0}^{N-1} \ell\left(X_{i|k}, \Pi_{i|k}\right)\\
    &\phantom{\leq} + \gamma \ell\left(X_{f|k}, \Pi_{f|k}\right) + \gamma F_{k+1}\left(G_{f|k}\left(X_{f|k}\right)\right)\\
    &\phantom{\leq} - F_{k}\left(X_{N|k}\right)\\
    &= -\ell\left(X_{0|k}, \Pi_{0|k}\right) + \gamma \ell\left(X_{f|k}, \Pi_{f|k}\right) \\
    &\phantom{=} + \gamma F_{k+1}\left(G_{f|k}\left(X_{f|k}\right)\right) - F_{k}\left(X_{N|k}\right)\\
    &\leq -\ell\left(X_{0|k}, \Pi_{0|k}\right) + \gamma \bar{\ell} + \gamma F_{k+1}\left(G_{f|k}\left(X_{f|k}\right)\right)\\
    &\phantom{\leq}  - F_{k}\left(X_{N|k}\right)
\end{align*}
where the last inequality follows from the definition of $\bar{\ell}$ in Corollary~\ref{cor:vf}.
Since $X_{N|k}\subseteq \gamma X_{f|k}$, according to the definition of the terminal cost
\begin{align*}
    &F_{k}\left(X_{N|k}\right)\\
    &\quad= \frac{\bar{\ell}}{1-\varrho}\left(M+(\lambda-1)\sigma(k+N)\right)\Psi_{\sigma(k)}\left(X_{N|k}\right)\\
    &\quad= \gamma \frac{\bar{\ell}}{1-\varrho}\left(M+(\lambda-1)\sigma(k+N)\right)\Psi_{\sigma(k)}\left(X_{f|k}\right)\\
    &\quad= \gamma F_{k}\left(X_{f|k}\right).
\end{align*}
Hence,
\begin{align*}
    \Delta V_k &\leq -\ell\left(X_{0|k}, \Pi_{0|k}\right)\\
    &\phantom{\leq} + \gamma \left(\bar{\ell} + F_{k+1}\left(G_{f|k}\left(X_{f|k}\right)\right) - F_{k}\left(X_{f|k}\right)\right)\\
    &\leq -\ell\left(X_{0|k}, \Pi_{0|k}\right) + \gamma \left(\bar{\ell} - \bar{\ell}W\left(k+N,X_{f|k}\right)\right)\\
    &\leq -\ell\left(X_{0|k}, \Pi_{0|k}\right)\\
    &\leq -s_8\left(\|x_{0|k}\|\right)
\end{align*}
where the second and third inequalities follow from Corollary~\ref{cor:vf}, and the last inequality from Assumption~\ref{ass:stage_kinf}.(i).
The fact that $V\left(k,x_{0|k},\mathbf{\Theta}_k\right)$ is monotonically decreasing with rate $s_8\left(\|x_{0|k}\|\right)$ is, in conjunction with the bounds of Assumption~\ref{ass:stage_kinf}.(ii), sufficient to conclude
that $V\left(\cdot,\cdot,\cdot\right)$ is a (time-varying) Lyapunov function.
Hence, asymptotic stability of the controlled system follows \cite[Theorem~2]{Aeyels1998}.
\qed

\paragraph*{Proof of Lemma~\ref{lem:cost}.}
For any $i\in\mathbb{N}_{[0,M-1]}$ let $p_i^X=(z_i,\alpha_i)$ and $p_i^\Pi=\left(u^{(1,1)}_i,\dots,u^{(t_i,q)}_i\right)$ be arbitrary but fixed tube parameters such that
$X_i=z_i\oplus\alpha_i S_i$ and $\Pi_i:X_i\times\Theta\rightarrow\mathbb{U}$ is an associated set-induced vertex controller according to \eqref{eq:gs_vertex_control}.
Since $\mathrm{rank}(Q)=n_\mathrm{x}$, $\exists a_i,b_i>0$ such that $\forall x\in X_i:\ a_i\|x\|\leq \|Qx\| \leq b_i\|x\|$ \cite[Corollary~II.8]{Lazar2010}.
Thus from \eqref{eq:stage_cost},
\begin{align*}
\ell\left(X_i,\Pi_i\right)
\geq \max_{x \in X_i} \|Q x\|
&\geq a_i \max_{x\in X_i} \|x\|
= a_i d_H^0\left(X_i\right).
\end{align*}
Hence Assumption~\ref{ass:stage_kinf}.(i) is satisfied with the global lower bound
$s_8(\xi) = \left(\min_{i\in \mathbb{N}_{[0,M-1]}} a_i\right) \xi$.
It is immediate that a lower bound on $V(x_0)$ is $s_{9}(\cdot)=s_8(\cdot)$.
The existence of a $\mathcal{K}_\infty$-upper bound $s_{10}(x_0)$ on $V(x_0)$ can then be shown proceeding as in \cite[Lemma~2]{Brunner2013}.
\qed
    
% ==============================================================================

\bibliographystyle{IEEEtran}
\bibliography{/home/jurre/tue/latex/bibtex/library_processed.bib}

% ==============================================================================

\end{document}